\documentclass[preprint]{aastex}
\title{The Origin of Jovian Planets in Protostellar Disks: The Role of Dead 
Zones}
\author{Soko Matsumura\thanks{soko@physics.mcmaster.ca} \ \& Ralph E. 
Pudritz\thanks{pudritz@physics.mcmaster.ca}}
\date{\hspace{0.2cm}Dept. of Physics \& Astronomy, McMaster University, 
Hamilton, ON L8S 4M1, Canada}
\begin{document}
%
\begin{abstract}
The final masses of Jovian planets are attained when the tidal torques that they exert on their surrounding protostellar disks are sufficient to open gaps in the face of disk viscosity, thereby shutting off any further accretion. 
In sufficiently well-ionized disks, the predominant form of disk viscosity 
originates from the Magneto-Rotational Instability (MRI) that drives 
hydromagnetic disk turbulence. 
In the region of sufficiently low ionization rate -- the so-called dead zone -- 
turbulence is damped and we show that lower mass planets will be formed.
We considered three ionization sources (X-rays, cosmic rays, and radioactive 
elements) and determined the size of a dead zone for the total ionization 
rate by using a radiative, hydrostatic equilibrium disk model developed by 
\citet{Chiang}.
We studied a range of surface mass density (\(\Sigma_{0}=10^3 - 10^5 \ {\rm g \ 
cm^{-2}}\)) and X-ray energy (\(kT_{x}=1 - 10\) keV).
We also compared the ionization rate of such a disk by X-rays with cosmic rays 
and find that the latter dominate X-rays in ionizing protostellar disks unless 
the X-ray energy is very high (\(5 - 10 \ {\rm keV}\)).
Among our major conclusions are that for typical conditions, dead zones 
encompass a region extending out to several AU -- the region in which 
terrestrial planets are found in our solar system.
Our results suggest that the division between low and high mass planets in 
exosolar planetary systems is a consequence of the presence of a dead zone in 
their natal protoplanetary disks.     
We also find that the extent of a dead zone is mainly dependent on the disk's 
surface mass density.
Our results provide further support for the idea that Jovian planets in 
exosolar systems must have migrated substantially inwards from their points of 
origin.
\end{abstract} 
%
\section{Introduction}
The discovery of exosolar Jovian-mass planets has stimulated intense efforts to 
understand how planets form \citep[e.g.][]{LBI, Wuchterl}.
The current theoretical models are based on two general types of physical 
processes -- disk instability or core accretion.
In the disk instability picture \citep[e.g.][]{Cameron, Boss97, Mayer}, 
gravitational instability in a sufficiently massive disk fragments it into the 
clumps out of which planets may form. 
This process may require as little as \(\sim 10^3\) years to form Jovian 
planets, but may not be able to explain sub-Jupiter mass planets \citep{Boss00}.
This is in stark contrast to the core accretion model \citep[e.g.][]{Mizuno, 
Pollack} wherein the coagulation of dust grains ultimately leads to the 
formation of planetesimals that collide with one another to form planetary 
cores.  
When a core reaches a critical mass in the range $ \sim 10 - 15 M_{\oplus} $, 
gas around the protoplanet quickly accretes onto it (``runaway accretion") 
forming a gas giant.
This process takes at least \(\sim 10^6-10^7\) years -- the average disk 
lifetime \citep[e.g.,][]{FM99}, and may not be able to explain massive Jovian 
planets.

Our paper focuses on an important aspect of Jovian planet formation that is 
common to both pictures. 
Regardless of whether proto-Jovian planets form through core accretion or 
gravitational instability, accretion onto a planet will continue until it is 
massive enough for the tidal torque that it exerts upon the surrounding disk to 
overcome the viscous disk torque which continues to fill-in any gap 
\citep[e.g.][]{LP}. 
Planets that are sufficiently massive to open gaps in the disks have final 
masses that depend upon the disk viscosity \citep{Nelson00}.  
In absence of viscosity, the tidal torque induced by the protoplanet opens a gap 
sooner thereby terminating the accretion earlier and leaving the planet with 
a smaller mass.

Thus the final mass of Jovian planets comes down to the question of the origin 
of the viscosity of protostellar disks at the time that significant 
protoplanetary cores are present within them.  If disks are turbulent, then 
turbulent viscosity is likely to be the major factor in determining these 
masses.                              
The Magneto-Rotational Instability (MRI) is the most promising source of 
turbulent disk viscosity \citep{BHin}.
Even in the presence of an initially very weak magnetic field, the MRI will 
generate a significantly magnetized disk.
This MRI viscosity requires good coupling between the gas and the magnetic 
field. 
Poor coupling -- which occurs when the gas is insufficiently ionized -- damps 
out the MRI and leads to a so-called ``dead zone" where the viscosity is 
effectively zero \citep{Gammie}.
The high column density for protoplanetary disks ensures that dead zones are 
expected to occur during planet formation \citep{Sano2}. 

Disk viscosity will not entirely vanish, even in the dead zone, however.  
The gravitational interaction between the fairly massive protoplanetary cores -- 
the forerunners of the Jovian planets -- and the disk generates density waves 
within the disk which can transport angular momentum and provide an effective 
``viscosity" in the disk.  
This disk-planet interaction leads to both planetary migration \citep{GT80} as 
well as evolution of the disk through the nonlinear damping and shocking of the 
waves \citep[][-- GR01]{Larson89, Goodman01}.      
The deposition of the angular momentum back into the disk as a consequence of 
this shock-induced wave damping is equivalent to the action of a viscous 
process, \citep{Spruit87, Larson89} with an equivalent dimensionless $\alpha$ 
parameter of the order $ \alpha_{damp} \simeq 10^{-4} - 10^{-3}$ (e.g. GR01).    
This viscosity parameter is one to two orders of magnitude smaller than is 
expected for MRI turbulence in the regions of the disk where the magnetic field 
is well coupled to the fluid.

We can readily compare the masses of planets that are predicted for each of 
these mechanisms of disk viscosity.    
For the case of a turbulent viscous disk \citep[e.g.][]{Bryden, Rafikov022}, the 
gap will open when the angular momentum transfer rate by tidal torque 
\(\dot{H_{T}}=0.23(M_{p}/M_{*})^2\Sigma a_{p}^4\Omega^2(a_{p}/h)^3\) 
exceeds that driven by viscous torque \(\dot{H_{\nu}}=3\pi\Sigma\nu 
a_{p}\Omega\).
Here \(\Omega\) is the Keplerian angular velocity and \(\nu\) is the viscosity. 
Expressing the viscosity in terms of a standard $\alpha$ parameter
\(\nu \sim \alpha_{turb}c_{s}h\) \citep{Shakura}, the ratio of a gap opening 
planetary mass \(M_{p}\) to a stellar mass \(M_{*}\) is
\begin{equation}
\frac{M_{p,turb}}{M_{*}} \lesssim 
\sqrt{40\alpha_{turb}\left(\frac{h}{a_{p}}\right)^5} \ .
\end{equation}
where the disk height and the radius of the planet's orbit 
are \(h\) and \(a_{p}\), respectively. 

In the case that disk viscosity is provided solely by damped density waves,
then the ratio of the planetary masses has been calculated by 
\citet{Rafikov022};   
\begin{equation}
\frac{M_{p,damp}}{M_{*}} \lesssim \frac{2}{3} 
\left(\frac{h}{a_{p}}\right)^3 {\rm min} \left[5.2 Q^{-5/7}, \ 3.8 \left(Q 
\frac{a_{p}}{h}\right)^{-5/13}\right] \,
\end{equation}
where the Toomre parameter is \(Q=\Omega c_{s}/(\pi G \Sigma)\), 
\(c_{s}\) is the sound speed, and \(\Sigma\) is the surface mass density.
Note that the gap opening condition for an inviscid disk by \citet{LP} 
corresponds to \(Q=1\) case.
For the standard disk model by \citet{Chiang}, we calculate that \(Q=54\) and 
\(h/a_{p} \sim 0.04\) at 1 AU, so that the planetary mass is very small.  

By taking the ratio of these two predicted planetary mass scales, we immediately see that the planetary masses in well coupled zones in which the MRI is active 
will be much larger than in the dead zone for which only the $\alpha_{damp}$ 
value pertains.  
As an example, at 1 AU, and for a Toomre Q value of 54 calculated above, the 
mass of a Jovian planet in a well-coupled zone dictated by MRI turbulence 
$M_{p,turb}$ will exceed the mass that a Jovian protoplanet will attain in a 
dead zone through density wave damping $M_{p,damp}$ provided that 
$\alpha_{turb}$ exceeds a minimum value of;    
\begin{equation}
\alpha_{turb} > \alpha_{min} \equiv {\rm min} \left[1.0 \times 10^{-3} {h \over 
a_p}, \ 7.4 \times 10^{-3} \left({h \over a_p}\right)^{23/13} \right]    
\end{equation}
which is less than $10^{-4}$ for the aspect ratio at 1 AU (\(h/a_{p} \sim 
0.04\)). 
Numerical simulations of the MRI instability show that $\alpha_{turb}$  
is much larger than this -- by two orders of magnitude or more.  Thus, a 
distinct jump in planetary masses -- by a factor that is similar to the ratio of the masses of the Jovian to terrestrial planets -- is expected at the disk 
radius that determines the maximum radial extent of the dead zone.        

In this paper, we perform a detailed investigation in order to calculate 
precisely where the dead zone occurs in well constrained protostellar disk 
models. 
This allows us to determine where in the disk that this ``jump" in planetary 
masses is likely to occur.   
In our second paper (to be submitted), we calculate precisely what
these planetary masses are. 
We proceed by carefully computing the ionization balance within
the protostellar disk, since it is this that determines the coupling
of the magnetic field to the disk and hence, where the MRI induced turbulence 
can be sustained. 
The main sources of ionization in a circumstellar disk 
are X-rays from the stellar plasma (probably produced by magnetospheric 
reconnection events) 
as well as cosmic rays from the outside of the stellar system.
\citet{Glassgold} showed that X-ray ionization dominates cosmic ray ionization 
out to \(\simeq 1000\) AU in an optically thin disk.
We compared these ionization sources for the self-consistent disk model 
developed by \citet{Chiang}.
We include radioactive elements as a global and constant ionization source.
We used the disk model developed by \citet{Chiang} primarily because it 
reproduces the spectral energy distributions (SEDs) of two T-Tauri and three 
Herbig Ae stars extremely well.

Our major finding is that dead zones typically encompass the region where 
terrestrial planets are found in our solar system (\(\lesssim 2.0 - 3.3\) AU).
The physical definition of a dead zone is reviewed in \S 2.
We introduce the disk model of \citet{Chiang} and elucidate the connection 
between the disk model and the X-ray ionization rate 
\citep[using][]{Glassgold97} in \S 3.
We state our results in \S 4 and apply them to the well studied case of AA Tau 
in \S 5. 
The important symbols used in this paper are summarized in Table \ref{table1}. 
\begin{table}
\caption{Some Important Symbols \label{table1}} 
\begin{center}
\begin{tabular}{|c|l|l|} \hline
\multicolumn{1}{|c|}{Symbol} &
\multicolumn{1}{c|}{Meaning} &
\multicolumn{1}{c|}{Reference} \\ \hline \hline
$M_{*}$ & Stellar mass &  \\
$R_{*}$ & Stellar radius & \\
$R_{\odot}$ & Solar radius & \\
$h$ & Pressure scale height of the disk & \\
$H$ & Surface disk height & \\
$a$ & Disk radius & \\ \hline
$Q$ & Toomre's Q parameter: $Q=\Omega c_{s}/(\pi G \Sigma)$ & eq. (2) \\
$c_{s}$ & Sound speed: $c_{s}=\sqrt{kT/(\mu m_{H})}$ & \S 1, 2 \\
$\nu$ & Viscosity & \S 1 \\
$\alpha_{turb}$ & $\alpha$ parameter of \citet{Shakura} & \S 1 \\
$V_{A}$ & Alfv\'en speed: $V_{A}=B/\sqrt{4 \pi \rho}$ & \S 2 \\
$\eta$ & Magnetic diffusivity & eq. (5) \\
$R_{eM}$ & Magnetic Reynolds number & eq. (4), (6), (7) \\
$x_{e}$ & Electron fraction & eq. (8), (9) \\
$x_{M}$ & Metal fraction & eq. (8) \\
$\beta$ & Recombination coefficient & eq. (9) \\
$\zeta$ & Ionization rate & eq. (8), (9), \S 3 \\ \hline
$\zeta_{tot}$ & Net ionization rate: $\zeta_{tot}=\zeta_{x} + \zeta_{CR} + 
\zeta_{RA}$ & eq. (10) \\
$\zeta_{x}$ & X-ray ionization rate & eq. (11) \\
$\zeta_{CR}$ & Cosmic ray ionization rate & eq. (16) \\
$\zeta_{RA}$ & Radioactive elements ionization rate & \S 3 \\
$\sigma$ & Total photoelectric absorption cross section & eq. (14) \\
$\tau$ & Optical depth & eq. (13) \\
$d$ & Distance from the X-ray source to the disk & eq. (11) \\
$n$ & Number density of neutral atoms & \S 2, eq. (17) \\
$N_{H}$ & Surface number density along the path of X-rays & eq. (13), (15) \\
$N_{\bot}$ & Vertical surface number density: 
$N_{\bot}=\int_{z}^{\infty}n(a,z^{\prime})dz^{\prime}$ & eq. (15) \\
$\chi$ & Vertical surface mass density: 
$\chi=\int_{z}^{\infty}\rho(a,z^{\prime})dz^{\prime}$ & eq. (16) \\
$\alpha$ & Grazing angle from the central star & eq. (21) \\
$\alpha^{\prime}$ & Grazing angle from the X-ray source at $(12 R_{\odot}, 12 
R_{\odot})$ & eq. (15), (21) \\ 
\hline
\end{tabular}
\end{center}
\end{table}
\begin{table}
\caption{Parameters used in the models. \label{table2}} 
\begin{center}
\begin{tabular}{|c|c|c|c|c|c|l|} \hline
\multicolumn{1}{|c|}{Fig.} & 
\multicolumn{1}{c|}{$R_{eM, crit}$} & 
\multicolumn{1}{c|}{$\alpha_{turb}$} & 
\multicolumn{1}{c|}{$L_{x} {\rm [erg \ s^{-1}]}$} & 
\multicolumn{1}{c|}{$kT_{x} {\rm [keV]}$} & 
\multicolumn{1}{c|}{$\Sigma_{0} {\rm [g \ cm^{-2}]}$} & 
\multicolumn{1}{c|}{$\zeta {\rm [s^{-1}]}$} \\ \hline \hline
\ref{plot2} & $1$ & $1, \ 0.1, \ 0.01$ & $10^{29}$ & $1$ & $10^3$ & 
$\zeta=\zeta_{x}$\\
\ref{plot3} & $1$ & $1, \ 0.1, \ 0.01$ & No X-rays & No X-rays & $10^3$ & 
$\zeta=\zeta_{CR}$ \\
\ref{plot4} & $1$ & $1, \ 0.1, \ 0.01$ & $10^{29}$ & $1$ & $10^3$ & 
$\zeta=\zeta_{tot}$ \\
\hline
\ref{plot5} & $1$ & $0.01$ & $10^{29}$ & $1, \ 2, \ 3, \ 5, \ 10$ & $10^3, \ 
5\times 10^3, \ 10^4, \ 6\times 10^4, \ 10^5$ & $\zeta=\zeta_{x}$ \\
\ref{plot6} & $1$ & $0.01$ & $10^{29}$ & $1, \ 2, \ 3, \ 5, \ 10$ & $10^3, \ 
5\times 10^3, \ 10^4, \ 6\times 10^4, \ 10^5$ & $\zeta=\zeta_{tot}$ \\
\ref{plot7} & $1$ & $0.01$ & $10^{29}$ & $1, \ 2, \ 3, \ 5, \ 10$ & $10^3, \ 
5\times 10^3, \ 10^4, \ 6\times 10^4, \ 10^5$ & $\zeta=\zeta_{x}$ \\
\ref{plot8} & $1$ & $0.01$ & $10^{29}$ & $1, \ 2, \ 3, \ 5, \ 10$ & $10^3, \ 
5\times 10^3, \ 10^4, \ 6\times 10^4, \ 10^5$ & $\zeta=\zeta_{tot}$ \\
\hline
\ref{plot9} & $1$ & $1, \ 0.1, \ 0.01$ & $0.439\times 10^{30}$ & $1.21$ & 
$1.5\times 10^3$ & $\zeta=\zeta_{x}, \  \zeta_{CR}$ \\
\ref{plot10} & $1$ & $1, \ 0.1, \ 0.01$ & $0.439\times 10^{30}$ & $1.21$ & 
$1.5\times 10^3$ & $\zeta=\zeta_{tot}$ \\
\hline
\end{tabular}
\end{center}
\end{table}
\begin{table}
\caption{Dead zone radii for the star with the X-ray luminosity of $10^{29} {\rm erg \ s^{-1}}$.  The range of dead zone radius corresponds to the range of surface mass density. \label{table3}} 
\begin{center}
\begin{tabular}{|c|c|c|c|c|c|} \hline
$(R_{eM, crit}, \alpha_{turb})$ & cosmic rays (CRs) & $kT_{x}=1$ keV & 
$kT_{x}=10$ keV & total(1keV+CR) & total(10keV+CR) \\ \hline \hline
$(1,1)$ & $0.70 - 6.5$ AU & $2.7 - 29$ AU & $0.24 - 3.7$ AU & $0.70 - 6.5$ AU & 
$0.24 - 3.7$ AU \\
$(1,0.01)$ & $1.3 - 8.6$ AU & $3.9 - 41$ AU & $0.66 - 6.7$ AU & $1.3 - 8.6$ AU & $0.66 - 6.7$ AU \\
$(100,1)$ & $2.7 - 13$ AU & $6.3 - 62$ AU & $1.6 - 13$ AU & $2.7 - 13$ AU & $1.5 - 12$ AU \\
$(100,0.01)$ & $5.8 - 21$ AU & $10 - 97$ AU & $3.6 - 28$ AU & $5.8 - 21$ AU & 
$3.6 - 20$ AU \\
\hline
\end{tabular}
\end{center}
\end{table}
\begin{table}
\centering
\caption{Input parameters for AA Tau \label{table4}} 
\begin{tabular}{|c|c|c|} \hline
Symbol & Meaning & Standard value \\ \hline \hline 
$M_{*}$ & stellar mass & $0.67M_{\odot}$ \\
$R_{*}$ & stellar radius & $2.1R_{\odot}$ \\ 
$T_{*}$ & stellar effective temperature & $4000$ K \\
$L_{x}$ & X-ray luminosity & $0.439\times 10^{30} \ {\rm erg \ s^{-1}}$ \\
$kT_{x}$ & X-ray energy & 1.21 keV \\ 
\hline 
$T^{{\rm iron}}_{sub}$ & iron sublimation temperature & $2000$ K \\
$T^{{\rm sil}}_{sub}$ & silicate sublimation temperature & $1500$ K \\ 
$T^{{\rm ice}}_{sub}$ & ice ${\rm H_{2}O}$ sublimation temperature & $150$ K \\ 
\hline
$\Sigma_{0}$ & surface mass density at 1 AU & $ 1.5 \times 10^3 \ {\rm g \ 
cm^{-2}}$ \\
$p$ & $- d \log \Sigma/d \log a$ & $1.5$ \\
$a_{i}$ & inner disk radius & $2 \ R_{*}$ \\
$a_{o}$ & outer disk radius & $8600 \ R_{*} = 250$ AU \\
$H/h$ & visible photosphere height / gas scale height & $3.8$ \\
$q_{i}$ & $- d \log N/d \log r$ in the interior disk & 3.5 \\
$q_{s}$ & $- d \log N/d \log r$ in the surface disks & 3.5 \\
$r_{{\rm max},i}$ & maximum grain radius in the interior disk & $1000 \ \mu m$ 
\\
$r_{{\rm max},s}$ & maximum grain radius in the surface disks & $1 \ \mu m$ \\
\hline
\end{tabular}
\end{table}
%
%
\section{Dead Zones in Disk Models}
A dead zone is the region in a disk where MRI turbulence is unsustainable due to 
poor coupling of the field to the disk.  
The absence of turbulence is what gives this region its negligible viscosity.
\citet{Gammie} showed that MRI turbulence will be damped at a 
physical scale of \(\lambda\) whenever the local growth rate of the 
MRI instability (\(\simeq V_{A}/\lambda\)) is balanced 
by the Ohmic diffusion at that scale (\(\simeq \eta/\lambda^2\)).
For complete damping at some radius of the disk, 
one requires that all the turbulence scales less than the 
local pressure scale height \(h(a)\) are damped; i.e. \(\lambda \lesssim h\).
By equating these two time-scales, one finds that the
MRI turbulence damps in regimes with a small magnetic Reynolds number:
\begin{equation}
R_{eM} = \frac{V_{A} h}{\eta} \lesssim 1 \ .
\end{equation}
Here, \(V_{A} \equiv B/\sqrt{4\pi \rho} \simeq \alpha_{turb}^{1/2} c_{s}\) 
is the Alfv\'en speed, \(\rho\) is the mass density, 
\(c_{s}=\sqrt{kT/\mu m_{H}}\) is the sound speed (\(k\) is the Boltzmann 
constant, \(T\) is the temperature, \(\mu\) is the mean molecular weight, and 
\(m_{H}\) is the mass of the hydrogen)
and \(\alpha_{turb}\) is the viscosity parameter that measures the strength of 
the turbulence \citep{Shakura}.
This formula is one of the most important in the paper because it is the 
condition that 
determines the dead zone region in a disk once the disk structure and ionization state are specified.   
The importance of the latter is readily established by noting that the  
Reynolds number in this equation is sensitive to the ionization of the disk, 
through its 
dependence on the diffusivity of the magnetic field,  
\(\eta\), which takes the form; 
\begin{equation}
\eta=\frac{234}{x_{e}}T^{1/2} \ {\rm cm^2 \ s^{-1}} \ ,
\end{equation}
where \(T\) is the disk temperature that is obtained from disk models 
\citep[either interior \(T_{i}\) or disk surface \(T_{ds}\) depending on the 
disk height we are interested in -- see \S 2.3, eq. (3) and (4) in][]{Chiang}. 
The electron fraction is \(x_{e}\equiv n_{e}/n\) where \(n_{e}\) and \(n\) 
are the number density of electrons and neutral atoms respectively.
The lower the electron fraction, the higher is the diffusivity coefficient for 
the field and hence the smaller is the magnetic Reynolds.  Thus, poorly ionized 
disks will tend to have lower values of $R_{eM}$, resulting in larger dead 
zones.

One of the recent numerical MRI disk simulations by \citet{Fleming} defined the 
magnetic Reynolds number as
\begin{equation}
R_{eM}^{\prime} = \frac{c_{s} h}{\eta} \ ,
\end{equation}
and determined the critical value \(R_{eM, crit}^{\prime}=100\) assuming the 
presence of a uniform vertical magnetic field.
Another recent analysis by \citet{Balbus01} suggested that the Hall effect 
should be included into the diffusion equation.
The Hall effect is due to the velocity difference between electrons and ions, 
and causes a small transverse potential difference.
The effect also diffuses the magnetic field just as the ohmic diffusion does.
It implies that the magnetic Reynolds number might be smaller than the case of 
\citet{Fleming}.

Summarizing these results, it appears that the critical value of 
magnetic Reynolds number \(R_{eM, crit}\) is somewhat uncertain and 
that there are a few different definitions for \(R_{eM}\).
Taking these uncertainties into account, we used the critical magnetic 
Reynolds number of 1 and 100 with the alpha parameter of 1, 0.1, and 0.01.
The extreme cases of \((R_{eM, crit}, \alpha_{turb})=\)(1, 1), (1, 0.01), (100, 
1), and (100, 0.01) correspond to \(R_{eM}^{\prime}=1, \ 10, \ 100, \ {\rm and} 
\ 1000\) respectively in the definition by \citet{Fleming}.
The following is the criterion we used to determine the dead zone: 
\begin{equation}
R_{eM} = \frac{V_{A} h}{\eta} \lesssim R_{eM, crit} \ .
\end{equation}
The dead zone is the region in the disk where the magnetic Reynolds number 
\(R_{eM}\) is smaller than some critical value \(R_{eM, crit} = 1 \ {\rm or} \ 
100\).  
This region is characterized as having no MRI turbulence.

The calculation of the needed electron fraction \(x_{e}\) (see eq. (5)) is well 
established, at least in principle.
Assuming steady state ionization balance, \citet{Oppenheimer} 
derived an equation for the electron fraction in the gas; namely,
\begin{equation}
x_{e}^3 + \frac{\beta_{t}}{\beta_{d}}x_{M}x_{e}^2 - \frac{\zeta}{\beta_{d} n} 
x_{e} - \frac{\zeta \beta_{t}}{\beta_{d} \beta_{r} n} x_{M} = 0 \ , 
\end{equation} 
where \(x_{M}\) is the metal fraction.
In our case, \(\zeta\) represents the ionization rate of X-rays, cosmic rays or 
radioactive elements, or total ionization rate of these three sources. 
Their mathematical forms are found in \S 3.
Three beta terms represent three kinds of recombination rate coefficients -- the dissociative recombination rate coefficient for electrons with molecular ions 
(\(\beta_{d}=2\times10^{-6}T^{-1/2} \ {\rm cm^3 \ s^{-1}}\)), the radiative 
recombination coefficient for electrons with metal ions 
(\(\beta_{r}=3\times10^{-11}T^{-1/2} \ {\rm cm^3 \ s^{-1}}\)) and the rate 
coefficient of charge transfer from molecular ions to metal atoms 
(\(\beta_{t}=3\times10^{-9} \ {\rm cm^3 \ s^{-1}}\)).

In this paper, we only take account of the two extreme cases -- the metal poor 
(\(x_{M}=0\)) and the metal dominant (\(x_{M}\gg x_{e}\)).
In both cases, the electron fraction is written as 
\begin{equation}
x_{e}=\sqrt{\frac{\zeta}{\beta n}} \ ,
\end{equation}
where the number density \(n\) depends on disk models (see eq. (17)) and 
\(\beta\) is a recombination rate coefficient.
In the metal poor case, only the recombination of electrons with molecular 
ions (dissociative recombination) is important and 
\(\beta=\beta_{d}=2\times10^{-6}T^{-1/2} \ {\rm cm^3 \ s^{-1}}\).
We use this coefficient throughout the paper except \S 5.
In the metal dominant case, only the recombination of electrons with heavy metal 
ions (radiative recombination) becomes important and the recombination rate 
coefficient is replaced by the radiative recombination coefficient 
\(\beta=\beta_{r}=3\times10^{-11}T^{-1/2} \ {\rm cm^3 \ s^{-1}}\) 
\citep{Fromang}.
The astrophysical significance of these two cases is related to the density 
regime of the protostellar disk.  
We compare the results for these two cases as applied to AA Tau, in  \S 5.
%
%
\section{Ionization Rate of a Self-Consistent Disk Model}
The previous section outlined precisely how the dead zone within a disk can be
calculated once a disk model and ionization rate can be specified.  
In this section, we review the three sources of ionization for protostellar 
disks -- X-rays from the central protostar, cosmic rays, and radioactive 
elements that are mixed with the disk material -- and then calculate their 
respective ionization rates in the context of our chosen disk model.
There is one caveat concerning cosmic ray ionization of protostellar disks -- cosmic rays are likely to be swept away from an region containing a turbulent MHD jet or outflow \citep[e.g.][]{Skilling, Cesarsky}.
Since jets are manifest in Class I and II Young Stellar Objects (YSOs), it is possible that the bulk of the ionization of these sources is achieved solely by their YSO X-ray fluxes.
For completeness, we show both sets of results -- ionization from purely X-ray as well as combined ionization.

The combined contribution of each of 
these ionization sources produces the total ionization rate which is 
their sum;  
\begin{equation}
\zeta_{tot}=\zeta_{x} + \zeta_{CR} + \zeta_{RA} \ .
\end{equation}
We summarize the calculation of each of these three rates below    
and then also compare the effects of X-rays with that of cosmic rays in our disk model calculations.
\subsection{Ionization Rates}
The X-ray ionization rate was investigated by \citet{Krolik} and further 
developed by \citet{Glassgold97}.
\citet{Glassgold} wrote the secondary electron contribution to the X-ray 
ionization rate \(\zeta_{x}\) as follows:
\begin{equation}
\zeta_{x} = \left[\left(\frac{L_{x}}{kT_{x} \ 4\pi d^2}\right)
\sigma(kT_{x})\right]\left(\frac{kT_{x}}{\Delta\epsilon}\right)J(\tau,x_{0}) \ ,
\end{equation}
where \(L_{x}\) is the observed X-ray luminosity, and \(\sigma(kT_{x})\) and 
\(\tau(kT_{x})\) are the total photoelectric absorption cross section and the 
optical depth at the energy \(E=kT_{x}\) respectively. 
The distance between the X-ray source and some point of the disk surface is 
denoted by ``d" and the energy to make an ion pair is \(\Delta \epsilon\).
In above equation, the first factor between the square brackets corresponds to 
the primary ionization rate, assuming the same energy \(E=kT_{x}\) for all 
primary electrons.
The second factor \(kT_{x}/\Delta \epsilon\) then reads as the number of 
secondary electrons produced by a photoelectron with the energy \(kT_{x}\).
The last factor \(J(\tau,x_{0})\) represents the attenuation of X-rays.
Using the dimensionless energy parameter \(x=E/kT_{x}\), the attenuation factor 
\(J(\tau,x_{0})\) is written as
\begin{equation}
J(\tau,x_{0}) = \int_{x_{0}}^{\infty} x^{-n} 
e^{-x-\tau\left(kT_{x}\right)x^{-n}} dx \ .
\end{equation}
The optical depth \(\tau(kT_{x})\) measures how opaque the disk is toward the 
X-ray radiation from the star and is written as 
\begin{equation}
\tau(kT_{x}) = N_{H} \ \sigma(kT_{x}) \ ,
\end{equation}
where \(\sigma(kT_{x})\) is 
\begin{equation}
\sigma(kT_{x})=\tilde{\sigma}_{0}\left(\frac{kT_{x}}{keV}\right)^{-n} \ .
\end{equation}
We assumed that heavy elements are depleted onto grains and used 
\(\tilde{\sigma}_{0}=8.50\times10^{-23} \ {\rm cm}^2\) and \(n=2.810\) 
\citep{Glassgold97}.
The surface number density \(N_{H}\) is measured along the radiation path from 
the X-ray source.
Letting \(\alpha^{\prime}\) be the angle between the radiation path and the 
radial axis, the surface number density is  
\begin{eqnarray}
N_{H} &=& \frac{N_{\bot}}{\sin\alpha^{\prime}} \nonumber \\
      &=& \frac{\int_{z}^{\infty} n(a,z^{\prime}) \ 
dz^{\prime}}{\sin\alpha^{\prime}} \ ,
\end{eqnarray}
where \(N_{\bot}\) is the vertical surface density and \(n(a,z)\) is the number 
density at the disk radius ``a" and the height ``z".
Note that \(n(a,z)\) and \(\alpha^{\prime}\) depend on the disk model (see eq. 
(17) and (21) respectively).

Usually, the ionization rate by cosmic rays is estimated as \(\zeta_{CR,0} 
\simeq 10^{-17} {\rm s^{-1}}\) \citep{Spitzer}, but it's not generally known to 
within better than an order of magnitude \citep{Glassgold}.
In some cases, however, the cosmic-ray ionization rate is well constrained. 
As an example, \citet{Tak} obtained the cosmic-ray ionization rate of 
\(2.6 \pm 1.8 \times 10^{-17} {\rm s^{-1}}\) through \({\rm H^{13}CO}\) 
submillimeter 
emission lines from massive protostars.
This value is in good agreement with Voyager/Pioneer data 
(\(4 \times 10^{-17} {\rm s^{-1}}\)).
Since the attenuation length for cosmic rays is \(\chi_{CR}\sim 96 \ {\rm g \ 
cm^{-2}}\) \citep{Umebayashi81}, we follow \citet{Sano2} and write the cosmic 
ray ionization rate as follows
\begin{equation}
\zeta_{CR}=\frac{\zeta_{CR,0}}{2}\left(\exp\left[-\frac{\chi 
(a,z)}{\chi_{CR}}\right] + \exp\left[-\frac{\Sigma (a) - \chi (a,z)}{\chi_{CR}} 
\right]\right) \ ,
\end{equation}
where \(\chi (a,z)\) is the vertical column density measured at some height 
\(z\) (\(\chi (a,z) = \int_{z}^{\infty}\rho (a,z^{\prime}) dz^{\prime}\)).
 
The radioactive elements in the disk is yet another source of ionization, though 
they usually have only a minor effect on the total ionization rate.
Following \citet{Umebayashi81}, we use the ionization rate by radioactive 
elements of \(\zeta_{RA}=6.9 \times 10^{-23} {\rm s^{-1}}\).

\subsection{Disk Model}
\citet{CG97}, following \citet{Kenyon}, developed a self-consistent passive disk 
model.
Their model has a two-layer structure -- a high temperature surface layer and a 
lower temperature interior.
Dust grains in the surface layer of the disk absorb the flux of ultra-violet 
(UV) photons from the central star.
Half of the emission from the grains in the 
surface layer escapes into the space and the remaining half heats the disk 
interior up.
The model assumes the vertical hydrostatic equilibrium so that the number 
density may be expressed as 
\begin{equation}
n(a,z) = n_{0}(a) \ e^{-\frac{z^2}{2h^2}} \ ,
\end{equation}
where the number density at the midplane \(n_{0}\) is 
\begin{equation}
n_{0}(a)=\Sigma (a) / (\sqrt{2\pi h(a)^2} \mu_{g}) \ .
\end{equation}
Here, \(\mu_{g}=3 \times 10^{-24} \ g\) is the mean molecular weight of the gas, 
\(\Sigma (a)\) is the surface density and \(h(a)\) is the pressure scale height 
\citep{Chiang}:
\begin{eqnarray}
\Sigma (a) &=& \Sigma_{0} \ \left(\frac{a}{{\rm AU}}\right)^{-3/2} \\
h(a) &=& \left(\frac{T_{i}}{T_{c}}\right)^{1/2} 
         \left(\frac{a}{R_{*}}\right)^{1/2} a \ ,
\end{eqnarray}
where \(\Sigma_{0}\) is the surface mass density at 1 AU, \(T_{c} \equiv GM_{*} 
\mu_{g}/kR_{*}\) is a virial temperature that measures the gravitational 
potential at the surface of the central star, and \(M_{*}\) and \(R_{*}\) are 
the stellar mass and radius respectively.
Note that the disk interior temperature \(T_{i}\) (the disk temperature below 
the pressure scale height ``h") decreases with a disk radius, so \(h(a)\) 
increases slightly weaker than the power of three halves (see eq. (20)) and 
\(n_{0}(a)\) decreases slightly weaker than the power of three (see eq. (18)). 
The pressure scale height \(h(a)\) is the height measured from the midplane to 
the interior disk, not the surface layer of the disk.

The ionization of this disk by cosmic rays and radioactive elements is 
straight-forward to calculate.
The X-ray ionization requires that we model the geometry of the source of the 
X-rays -- the central young stellar object.  
X-ray emission is thought to arise from the reconnection of large-scale 
magnetized loops, that are possibly but not necessarily related to the 
magnetopause radius of the stellar magnetic field and the inner edge of the 
disk.       
Accordingly, we place the X-ray source in this picture at a fiducial distance of
\(12 R_{\odot}\) above the disk and \(12 R_{\odot}\) away from the central star 
following \citet{Glassgold97}.  

\begin{figure}[t]
\includegraphics[width=10cm]{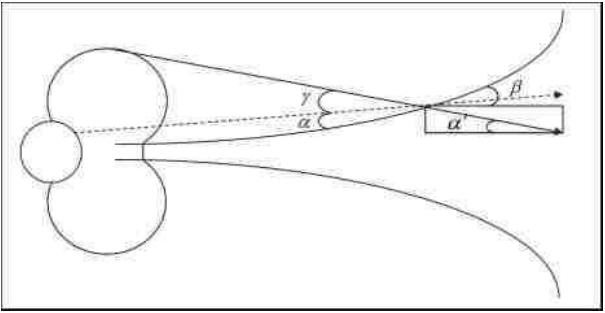}
\caption[The grazing angle]{A schematic figure of the cross section of the disk.  The solid arrow shows the X-rays from the magnetic field (the X-ray source is 
at (\(12 R_{\odot}, 12 R_{\odot}\))) and the dashed arrow shows the light from 
the central star.
Note that there is a relation of \(\alpha^{\prime}=\alpha-\beta+\gamma\) among 
angles (see eq. (21) and the text). \label{plot1}}
\end{figure}

The X-rays from this source graze the surface of the flaring disk and penetrate 
it to produce the X-ray induced ionization.  
It is convenient to relate this grazing angle, $\equiv \alpha^{\prime}$, to the 
grazing angle defined in the disk model of \citet{Chiang} \(\equiv \alpha\) 
\citep[see eq. (5) in][]{Chiang}.
Assuming that the average radiation from the central star originates at a 
distance \(R_{*}/2\) from the disk plane on the stellar surface 
\citep{Hartmann}, we write the grazing angle from the X-ray source (see eq. 
(15)) as
\begin{eqnarray}
\lefteqn{\alpha^{\prime} = \alpha - \beta + \gamma } \nonumber \\
& & \beta \equiv \tan^{-1} \left( \frac{d \ {\rm ln H}}{d \ {\rm ln 
a}}\frac{H}{a} \right) \nonumber \\
& & \gamma \equiv 
\tan^{-1}\left(\frac{H-\frac{1}{2}R_{*}}{a-\frac{\sqrt{3}}{2}R_{*}}\right)
    -\tan^{-1}\left(\frac{H-12R_{\odot}}{a-12R_{\odot}}\right) \ ,
\end{eqnarray}
where \(H\) is the height of the surface layer of the disk and defined as 
\(H/h\equiv 4\) in \citet{Chiang}.
Fig. \ref{plot1} is a schematic figure of this equation.
The grazing angle from the X-ray source, \(\alpha^{\prime}\) is the angle 
between the disk midplane and the solid arrow.
The first term of the righthand side of the equation, \(\alpha\) is the angle 
between the disk surface and the dashed arrow of the UV emission from the star. 
The second term \(\beta\) shows the flaring angle of the disk and the third term 
\(\gamma\) comes from the geometry of the X-ray source.
Note that \(\gamma=0\) corresponds to an X-ray source that is at the stellar 
surface.
 
We calculated the X-ray ionization rate (eq. (11)) by integrating the 
attenuation factor (eq. (12)) using MATLAB.  
Instead of an infinite upper limit of the energy, we used \(x=100\) for the 
upper limit of the integration.
For the lower limit, we used \(x=1 \ {\rm keV} / kT_{x}\) (\(E=1 \ {\rm keV}\)), 
following \citet{Glassgold97}.
The surface number density along the path of the radiation \(N_{H}\) found in 
the optical depth (eq. (13)) is calculated by using the grazing angle 
\(\alpha^{\prime}\) (eq. (21)) and the number density \(n(a,z)\) that is 
obtained from the disk model of \citet{Chiang} (see eq. (17) and (18)).
Note that this number density \(n(a,z)\) depends on the disk interior 
temperature \(T_{i}\) and the pressure scale height of the disk \(h\). 
Both of them are determined by solving the radiative equilibrium equations (3) 
and (4) in \citet{Chiang}.
%
%
\section{Results}
In this section, we study the extent of dead zones for various disks.
We show that X-rays and cosmic rays are comparable ionization sources and 
determine the size of dead zones from the total ionization rate.
Throughout this section, we assume the metal poor disk (set \(\beta = 
\beta_{d}\) in eq. (9)).
\subsection{Standard Disk Model}
First, we show the results of a fiducial model which uses the standard disk 
model of \citet{Chiang}, the critical magnetic Reynolds number \(R_{eM, 
crit}=1\) with \(\alpha_{turb}=1, \ 0.1, \ 0.01\), and when X-rays exist, the 
X-ray luminosity of \(L_{x}=10^{29} \  {\rm erg \ s^{-1}}\) with the temperature of \(kT_{x}=1\) keV.
We chose the standard X-ray luminosity of \(L_{x}=10^{29} {\rm erg \ s^{-1}}\) 
because typical young stars have X-ray luminosities \(L_{x} \sim 10^{28} - 
10^{30} {\rm erg \ s^{-1}}\). 
Though the standard disk of \citet{Chiang} has the surface mass density at 1 AU 
of \(\Sigma_{0}=10^3 \ {\rm g \ cm^{-2}}\), we don't call this minimum mass 
solar nebular model because these two models are qualitatively different from 
each other.
With the standard disk model of \citet{Chiang}, we calculated the surface number
density (eq. (15)) and/or the surface mass density (\(\chi\) in eq. (16)) to 
estimate the X-ray and/or cosmic ray ionization rate at each radius and height 
of the 2D disk.
All disk parameters are the same as in Table 1 of \citet{Chiang}.
Other parameters used in this subsection is summarized in Table \ref{table2} in 
this paper.
We determined the critical height of the dead zone at each radius of the disk 
from eq. (7).

\begin{figure}[t]
\includegraphics[width=9cm]{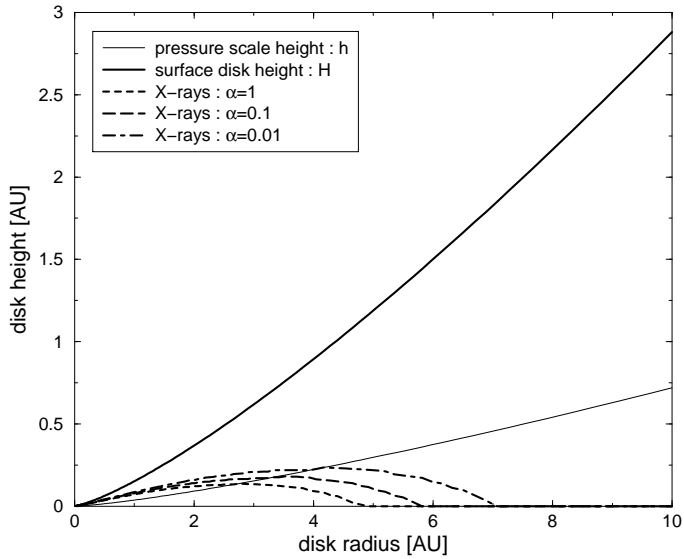}
\caption[The dead zone from X-rays]{The dead zone estimated by X-rays for the 
standard disk model of \citet{Chiang}.  We used \(L_{x}=10^{29} \ {\rm erg \ 
s^{-1}}\), \(kT_{x}=1 \ {\rm keV}\), \(R_{eM, crit}=1\) and \(\alpha_{turb}=1, \ 0.1, \ \& \ 0.01\).  X-ray dead zone stretches out to \(5 - 7\) AU. 
\label{plot2}}
\end{figure}

The size of dead zones arising solely from X-ray irradiation are plotted in Fig.
\ref{plot2}.
The figure shows the vertical cross section of the standard disk.
The region below the dashed, long-dashed and dot-dashed lines show the dead zones.
As the magnetic field becomes weaker (the alpha value changes \(\alpha_{turb}=1, \ 0.1, \ 0.01\)), the dead zone becomes larger.
The result follows from the fact that as the magnetic field becomes weak, the 
growth rate of the MRI instability is reduced with respect to the local damping 
rate (see eq. (4) and the text therein).

\begin{figure}[p]
\includegraphics[width=9cm]{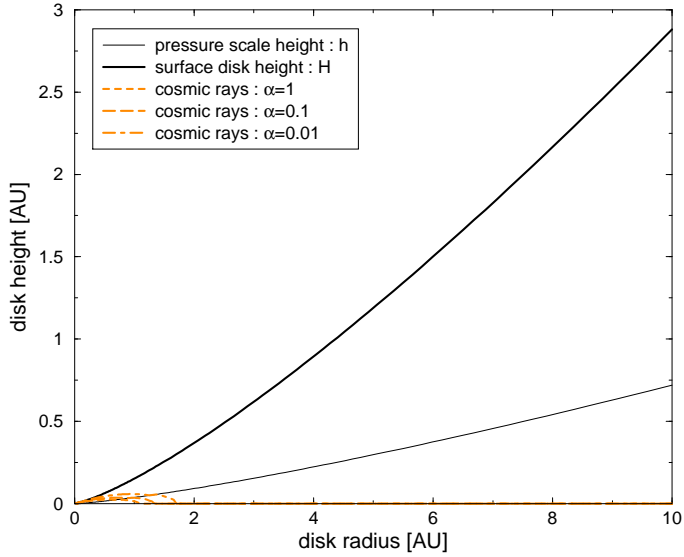}
\caption[The dead zone from cosmic rays]{The dead zone estimated by cosmic rays 
for the standard disk model of \citet{Chiang}.  We used \(R_{eM, crit}=1\) and 
\(\alpha_{turb}=1, \ 0.1, \ \& \ 0.01\). Cosmic ray dead zone stretches out to 
\(1 - 2\) AU. \label{plot3}}
\end{figure}

The equivalent result for cosmic rays is shown in Fig. \ref{plot3}.
In this case, we just replaced the ionization rate in electron fraction (eq. 
(9)) with the cosmic ray ionization rate calculated by eq. (16).
Again, we find the smaller dead zone as the magnetic field becomes stronger.

These two figures show that cosmic rays dominate 1 keV X-rays in ionizing a 
disk.
The dead zone stretches out to \(\sim 5 - 7\) AU in the case of illumination by 
X-rays and only to \(\sim 1 - 2\) AU in cosmic ray case. 

The dividing line between X-ray or cosmic ray ionized regions may be found by 
equating their ionization rates.
We call this resulting vertical scale height the R\"ontgen height.
Note that due to the higher number density, the X-ray ionization rate decreases 
toward the disk midplane as does the electron fraction, while the cosmic ray 
ionization rate is about the same everywhere in the disk.
X-ray ionization dominates cosmic ray ionization above the R\"ontgen height. 

\begin{figure}[t]
\includegraphics[width=9cm]{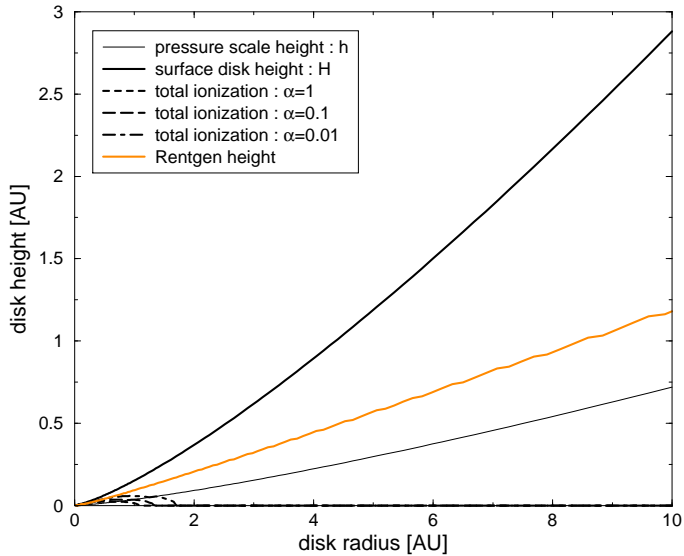}
\caption[The dead zone for the standard disk]{The dead zone determined by the 
total ionization rate for the standard disk model of \citet{Chiang}.  We used 
\(L_{x}=10^{29} \ {\rm erg \ s^{-1}}\), \(kT_{x}=1 \ {\rm keV}\), \(R_{eM, 
crit}=1\) and \(\alpha_{turb}=1, \ 0.1, \ \& \ 0.01\). Also shown is the 
R\"ontgen height.  X-rays dominate cosmic rays in ionizing a disk only at the 
surface layer, so the dead zone estimated from the total ionization rate has the same size as that from cosmic rays. \label{plot4}}
\end{figure}

Fig. \ref{plot4} shows the dead zones estimated by the total ionization rate.
Again, we took a surface mass density at 1 AU of \(\Sigma_{0}=10^3 \ {\rm g \ 
cm^{-2}}\), the dead zone criterion \(R_{eM}\leq 1\), an X-ray luminosity of 
\(L_{x}=10^{29} {\rm erg \ s^{-1}}\), and an X-ray energy of \(kT_{x}=1\) keV. 
The R\"ontgen height suggests that X-rays dominate cosmic rays only in the surface layer -- where the disk is optically thin.
Comparing Fig. \ref{plot3} with \ref{plot4}, it is apparent that the size of 
dead zones are determined by cosmic rays for the standard disk case.

There are two characteristic dead zone radii.
One of them is the midplane radius where the dead zone surface cuts the disk 
midplane.
The other one is the fiducial radius where the dead zone surface crosses 
the pressure scale height of the interior disk.
We take this latter case to be our ``fiducial" value because planets are likely 
to form in the interior disk that has much higher density (the order of \(\sim 
10^3\)) and larger particles (\(0.01 - 1000 \mu m\) instead of \(0.01 - 1 \mu 
m\)) compared to the surface layers.
\subsection{Sensitivity of Dead Zone Radius to Model Parameters}
Next, we compare cases of different parameters (\(\Sigma_{0}, kT_{x}\)) with one another.
This is done to reveal which parameter has the largest effects on the size of 
dead zone.
We also present results for two cases; one for only X-ray induced ionization, 
and the other for ionization arising from the total ionization rate.  
We make this distinction in order that the contribution of X-rays can be clearly discerned (cf. discussion at the beginning of \S 5).      
Throughout this subsection, we set \(R_{eM, crit}=1\) and \(\alpha_{turb}=0.01\)
(see Table \ref{table2}).
These are the conditions suggested by, for example, \citet{Gammie}.

Fig. \ref{plot5} is the plot of the fiducial dead zone radius estimated by X-ray alone as a function of the X-ray energy (\(kT_{x}=1, \ 2, \ 3, \ 5, \ {\rm and} 
\ 10\) keV).
For the X-ray ionization calculation, we used the typical X-ray luminosity of 
\(L_{x}=10^{29} {\rm erg \ s^{-1}}\).  
We include the minimum mass solar nebula (\(\Sigma_{0}=10^3 \ {\rm g \ 
cm^{-2}}\)) as well as a very massive disk of \(\Sigma_{0}=10^5 \ {\rm g \ 
cm^{-2}}\) suggested by \citet{Murray} for Jupiter to migrate from 5 AU to \(< 
1\) AU. 
It is apparent that the sizes of dead zones decrease as the X-ray energy 
increases and/or the disk surface mass density decreases.

Note that the figure depends on the choice of the critical magnetic Reynolds 
number \(R_{eM, crit}\), the alpha parameter \(\alpha_{turb}\), and the X-ray 
luminosity \(L_{x}\).
For example, if we increase the X-ray luminosity by two orders of magnitude 
(\(L_{x}=10^{31} {\rm erg \ s^{-1}}\)) in Fig. \ref{plot5}, then the resulting 
curves are pushed down by the factor of 10.
This is because of the definition of the magnetic Reynolds number \(R_{eM} = 
\sqrt{\alpha_{turb} \zeta} \ [c_{s}h/(234\sqrt{\beta n T})]\).
Here, the bracket term is determined by a disk model (i.e. independent of the 
X-ray energy or luminosity) so that the magnetic Reynolds number is changed 
depending on \(\alpha_{turb}\) and \(\zeta\).
For X-rays, the ionization rate is written as \(\zeta_{x} = L_{x} \ 
[\sigma(kT_{x})/(4 \pi d^2 \Delta \epsilon) \ J(\tau, x_{0})]\) where the 
luminosity changes the result by two orders of magnitude in our case (we used 
\(L_{x}=10^{29} {\rm erg \ s^{-1}}\) and \(10^{31} {\rm erg \ s^{-1}}\)).

\begin{figure}[p]
\includegraphics[width=9cm]{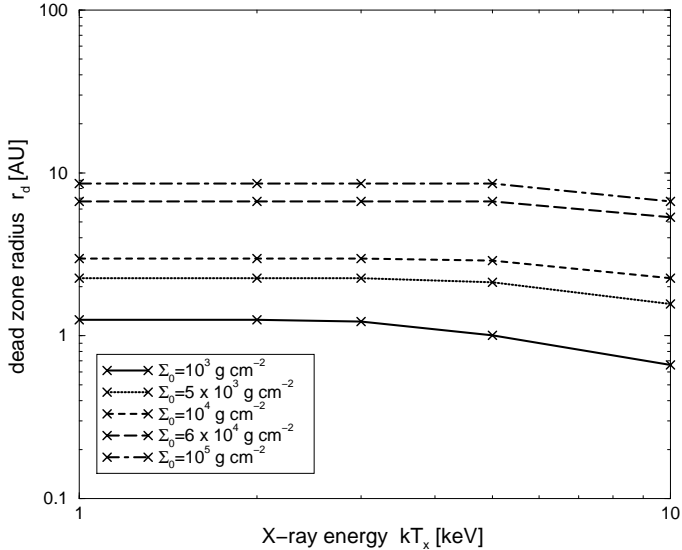}
\caption[Dead zone vs X-ray energy 8]{The fiducial dead zone radius as a 
function of the X-ray energy for a source luminosity \(L_{x}=10^{29} {\rm erg \ 
s^{-1}}\). We chose \(R_{eM, crit}=1\) and \(\alpha_{turb}=0.01\) and estimated 
the dead zone sizes by total ionization rates.  Cosmic rays dominate low energy X-rays in ionizing a disk. \label{plot6}}
\end{figure}

The results for the dead zone radius for a disk undergoing the total ionization 
rate is shown in Fig. \ref{plot6}. 
It shows that the radii of dead zones are almost constant for X-ray energy of 
\(kT_{x}=1 - 3 \ {\rm keV}\) and coincide with the results of Fig. \ref{plot5} 
in the higher X-ray energy region.

These figures also show the obvious point that the size of dead zones 
is not affected by low energy X-rays (\(kT_{x} \sim 1 - 3 \ {\rm keV}\)) 
and that the dead zone gets smaller as the surface mass density decreases.
The decrease of the surface mass density (from upper lines to lower ones) leads 
to a smaller optical depth \(\tau\) and a larger ionization rate \(\zeta\), so 
that the dead zone becomes smaller. 

We summarize the effect of changing parameters. 
If the X-ray luminosity is larger (\(10^{31} {\rm erg \ s^{-1}}\) rather than 
\(10^{29} {\rm erg \ s^{-1}}\)), X-rays penetrate deeper in the disk and 
therefore the dead zone becomes smaller.
If the critical magnetic Reynolds number is larger (100 rather than 1), a very 
small magnetic field diffusivity can destroy the turbulence -- the viscosity 
becomes negligible and therefore the dead zone tends to be larger.
If the alpha parameter is smaller (0.01 rather than 1), then the magnetic field 
is weaker, turbulent viscosity is smaller and the dead zone becomes larger.
In short, to have a smaller dead zone, a disk has to have a smaller surface mass density \(\Sigma_{0}\), a larger X-ray luminosity \(L_{x}\) and a larger 
\(\alpha_{turb}\) value as well as a smaller critical Reynolds number 
\(R_{eM,crit}\).

To draw the reader's attention to our major result, we present Fig. \ref{plot7} 
and \ref{plot8} and study the effect of the surface mass density at 1 AU 
\(\Sigma_{0}\).
Fig. \ref{plot7} uses the same data sets as Fig. \ref{plot5}, but here we plot 
the fiducial dead zone radius as a function of the surface mass density at 1 AU 
(\(\Sigma_{0}=10^3, \ 5\times10^3, \ 10^4, \ 6\times10^4, \ {\rm and} \ 10^5 \ 
{\rm g \ cm^{-2}}\)) instead of the X-ray energy.
It is apparent that if cosmic rays penetrate disks without being swept away, they dominate X-rays in ionizing a disk except for \(kT_{x} \sim 5 - 10 \ {\rm keV}\).

Fig. \ref{plot8} shows the equivalent result for the total ionization rate and 
uses the same data sets as Fig. \ref{plot6}.
The X-rays have a rather small effect on the size of dead zones even when they 
have high energy (\(kT_{x} \sim 5 - 10 \ {\rm keV}\)).
The most striking result is that dead zone radii vary by at most an order of 
magnitude, from \(1 - 10\) AU for a range of two orders of magnitude in disk 
surface density and an order of magnitude in the X-ray energy.  
The robustness of this result indicates that the region of Jovian-mass planet 
formation is rather similar for a wide variety of protostellar disk systems.

We can see that the differences between lines in Fig. \ref{plot7}, \ref{plot8} 
are smaller than those in Fig. \ref{plot5}, \ref{plot6}.
It appears that the size of the dead zone is most sensitive to the surface mass 
density and, to a lesser extent, the X-ray energy. 
These figures also show that the dead zone produced by cosmic rays is usually 
smaller than that by X-rays.
X-rays dominate cosmic rays in ionizing the circumstellar disk only when the 
X-ray source has a high energy (\(kT_{x} \sim 5 - 10 \ {\rm keV}\)). 

Table \ref{table3} shows the size range of dead zones for four extreme cases -- 
\((R_{eM, crit}, \alpha_{turb})=(1, 1)\), (1, 0.01), (100, 1), and (100, 0.01).
The minimum and maximum dead zone radii correspond to the minimum and maximum 
surface mass density at 1 AU.
It is apparent that the dead zone is sensitive not only to the surface mass 
density or X-ray energy, but also to the value of the magnetic Reynolds number.
The important point of this table is that the dead zone radii for total 
ionization rate are about the same as the region of terrestrial planets in our 
solar system except the one extreme case.
For this extreme case of \((R_{eM, crit}, \alpha_{turb})=(100, 0.01)\), which is 
equivalent to \(R_{eM, crit}^{\prime}=1000\) or \((R_{eM, crit}, 
\alpha_{turb})=(1, 10^{-6})\), the dead zone gets pushed out to 5.8 AU.

Finally, we note that the data presented in Figures \ref{plot5}, \ref{plot7} and 
\ref{plot8} corresponding to the variation of dead zone radii can be well fit by 
approximate linear curves.  
For completeness, we present these fits in Appendix A - Table \ref{table5}, 
\ref{table6}, \ref{table7}, and \ref{table8}.  
%
%
%
\section{Application to the case of AA Tau}
We now specialize our model to an observed protostellar disk -- AA Tau.
We consider this source because of the excellent fit of the Chiang et al. model 
to the data from this system.
The stellar and disk parameters that we used are listed in Table \ref{table4} 
and are taken from \citet{Chiang} other than X-ray parameters.
The X-ray luminosity \(L_{x}\) and the X-ray energy \(kT_{x}\) are taken from 
\citet{Neuhauser}.
For the dead zone criterion, we used \(R_{eM, crit}=1\) with \(\alpha_{turb}=1, 
\ 0.1, \ 0.01\).
These parameters are summarized in Table \ref{table2}.

\begin{figure}[p]
\includegraphics[width=9cm]{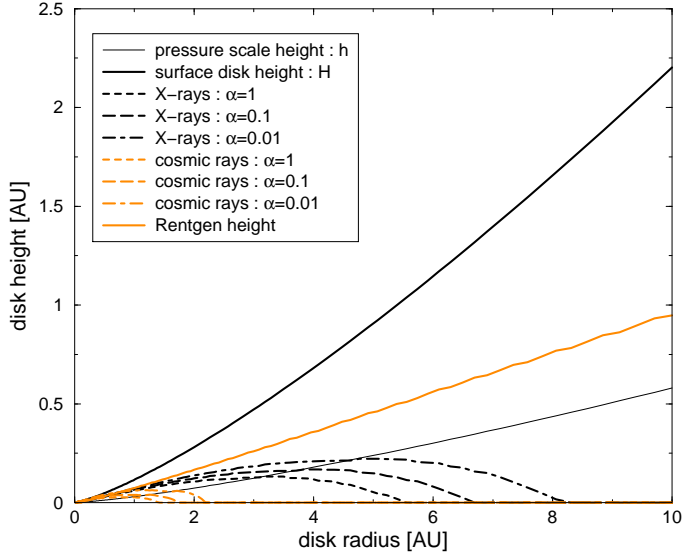}
\caption[Dead zone for AA Tau]{The dead zone of AA Tau estimated from X-rays is 
compared with that predicted from cosmic rays.  We used \(L_{x}=0.439 \times 
10^{30} \ {\rm erg \ s^{-1}}\), \(kT_{x}=1.21 \ {\rm keV}\), \(R_{eM, crit}=1\),
and \(\alpha_{turb}=1, \ 0.1, \ {\rm and} \ 0.01\).  Also shown is the R\"ontgen
height.  Cosmic rays dominate X-rays in ionization almost throughout the disk. 
\label{plot9}}
\end{figure}

Fig. \ref{plot9} is the estimated dead zone for AA Tau. 
It shows that X-ray ionization dominates cosmic ray ionization only in the 
surface layer where the optical depth is very small (see R\"ontgen height).
The dead zone becomes smaller as the magnetic field gets stronger because of the 
stronger viscous torque that it exerts.
In all cases the dead zones by cosmic rays are smaller than those produced by 
X-rays.

\begin{figure}[p]
\includegraphics[width=9cm]{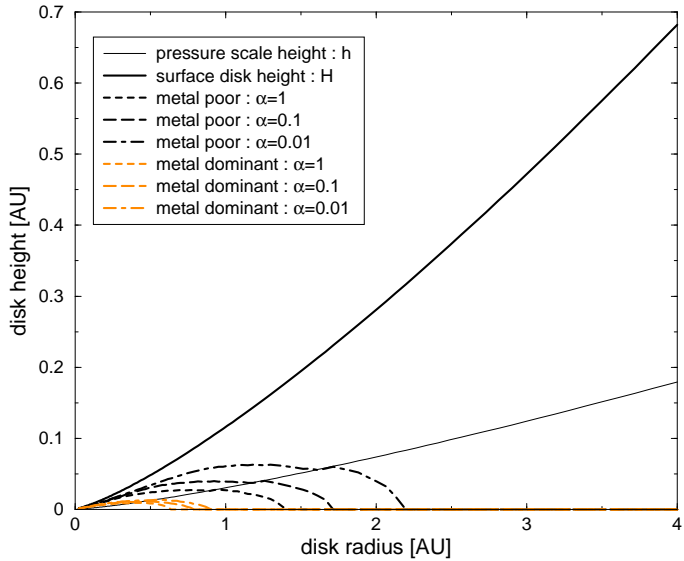}
\caption[Metal dominant and metal poor cases for AA Tau]{The dead zone of AA Tau
estimated both for a metal poor and a metal dominant disk.  We used \(L_{x}=0.439 \times 10^{30} \ {\rm erg \ s^{-1}}\), \(kT_{x}=1.21 \ {\rm keV}\), \(R_{eM, crit}=1\), and \(\alpha_{turb}=1, \ 0.1, \ {\rm and} \ 0.01\). The presence of metals makes the dead zone smaller. \label{plot10}}
\end{figure}

If the disk of AA Tau is metal dominant, the dead zone gets smaller.
Fig. \ref{plot10} shows the dead zones estimated by total ionization rate for 
both metal poor and metal dominant cases.
Since metal ions recombine with electrons slowly, the electron fraction for 
metal dominant case is larger (compare \(\beta_{d}\) with \(\beta_{r}\)) -- 
\(R_{eM}\) tends to be larger -- and hence the dead zone is smaller. 
Note that the recombination coefficient of metals is \(\sim 5\) orders of 
magnitude smaller than that of molecular ions.
The ``real" dead zone would be calculated by solving eq. (8) directly and would 
be located somewhere between these two extremes.
%
%
\section{Discussion and Conclusions}

We calculated the dead zones for a variety of disk models, as well as the 
specific case of a Class II source, AA Tau.
This kind of calculation has been done by several authors 
\citep[e.g.,][]{Gammie, Glassgold99, Sano2, Fromang}, but all of them use either 
the minimum mass solar nebula model developed by \citet{Hayashi} 
\citep[e.g.,][]{Gammie, Glassgold99, Sano2} or the \(\alpha\) disk model 
developed by \citet{Shakura} \citep{Fromang}.
All our disk models are obtained by using the self-consistent passive disk model 
of \citet{Chiang}. 
To calculate X-ray ionization rates, we followed the method used by 
\citet{Glassgold}, while for cosmic rays, we used the method adopted by 
\citet{Sano2}.
When the ionization rate becomes sufficiently low so that the magnetic Reynolds 
number is less than some critical value (1 or 100), we determined the size of 
the dead zone.
We compared the result by X-rays with cosmic rays and also obtained the size of 
the dead zone determined by the total ionization rate.

The fact that our dead zones extend as far as the region of terrestrial planet 
formation in our solar system is very important.
In the accretion picture, considerable gas accretion must occur onto a 
sufficiently massive protoplanetary core.
This must inevitably take place in the turbulent region of the disk beyond the 
dead zone.
The major implication of our work is that the division between 
Jovian and sub-Jovian or terrestrial planets may be due to the presence of a 
dead zone.
We explore the significance of these results for predicting planetary masses in 
a forthcoming paper. 
Our basic results are as follows: 
\\

1. Our major finding is that the typical dead zone encompasses, 
in physical scale, the terrestrial planets in our solar system.

A typical dead zone size estimated from the total ionization rate is \(\sim 0.24 - 2.7\) AU for the fiducial disk.
For an extreme case of \((R_{eM, crit}, \alpha_{turb})=(100, 0.01)\) (\(R_{eM, 
crit}^{\prime}=1000\) or \((R_{eM, crit}, \alpha_{turb})=(1, 10^{-6})\) 
equivalently), the dead zone extends out to \(\sim 5.8\) AU. 
These results suggest that the observed exosolar Jovian-mass planets must have 
migrated to their observed positions from points of origin farther out in disk 
radius, beyond these dead zone radii. 
 
2. X-rays dominate cosmic rays in ionizing a disk only in the surface of the 
disk for most cases.

\citet{Glassgold} noted that X-ray ionization dominates cosmic ray ionization 
out to \(\simeq 1000\) AU.
We found that the disks are usually too optically thick for this to be generally 
true.
X-rays only dominate cosmic rays in the surface layer of the disk. 
Cosmic rays determine the size of the dead zone under typical conditions 
(compare, for example, Fig. \ref{plot3} with Fig. \ref{plot4}). 

3. Sufficiently high energy X-rays could dominate cosmic rays in ionizing a 
disk.

The X-rays could dominate the cosmic rays if the X-ray energy is high \(kT_{x}=5 
- 10 {\rm keV}\) (compare Fig. \ref{plot7} with \ref{plot8}). 
This is much higher energy for most observed sources however.

4. The size of a dead zone is sensitive to the disk surface mass density 
\(\Sigma_{0}\) and the X-ray energy \(kT_{x}\).

We found that the size of the dead zone is most sensitive to the surface mass 
density, and to a lesser extent, the X-ray energy (see Fig. \ref{plot5} \& 
\ref{plot6} and Fig. \ref{plot7} \& \ref{plot8}).
The X-ray luminosity, the critical magnetic Reynolds number, the alpha viscous 
parameter however, have a significant effect on the size of a dead zone. 

5. There is a power law relation between the size of a dead zone estimated by 
X-rays alone and \(\Sigma_{0}\) or \(kT_{x}\) (see Appendix).

The dead zone radius has power-law relations with both the surface mass density 
and the X-ray energy.
The dead zone radii estimated by X-rays obey \(a_{d} \propto \Sigma_{0}^{0.44} - \Sigma_{0}^{0.59}\) and \(a_{d} \propto kT_{x}^{-0.46} - kT_{x}^{-1.1}\), while those estimated by cosmic rays obey \(a_{d} \propto \Sigma_{0}^{0.28}  - 
\Sigma_{0}^{0.49}\).
The dead zone radii determined by the total ionization rate has the relation of 
\(a_{d} \propto \Sigma_{0}^{0.28}  - \Sigma_{0}^{0.59}\). 

We thank Norm Murray, Joe Weingartuer and Steve Balbus for stimulating 
conversations on these topics.
We also thank an anonymous referee for useful review of the manuscript.
SM is supported by McMaster University, while
REP is supported by a grant from the National Science and Engineering Research 
Council of Canada (NSERC).
\bibliographystyle{apj}
\bibliography{REF}
\clearpage
\begin{appendix}
\section*{\Large{Appendix A}}
\section*{X-rays and Cosmic rays}
We supply tables of X-ray and cosmic ray power-law relations separately for 
interested readers.
These approximate relations are all obtained by using the least square fitting.

\begin{figure}[p]
\includegraphics[width=9cm]{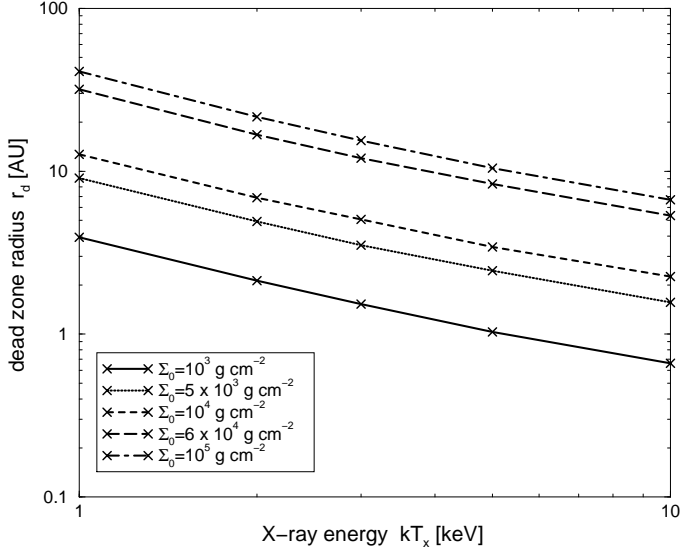}
\caption[Dead zone vs X-ray energy 4]{The fiducial dead zone radius as a 
function of the X-ray energy for a source luminosity \(L_{x}=10^{29} {\rm erg \ 
s^{-1}}\). We chose \(R_{eM, crit}=1\) and \(\alpha_{turb}=0.01\) and estimated 
the dead zone sizes by X-rays alone. Higher X-ray energy and smaller surface mass density lead to the smaller dead zone.\label{plot5}}
\end{figure}

Table \ref{table5} shows the fitting parameters for the dead zone radius by 
X-rays as a function of X-ray energy \(kT_{x}\). 
For the same choice of the critical magnetic Reynolds number \(R_{eM, crit}\) 
and the alpha parameter \(\alpha_{turb}\), it is apparent that the curves show 
roughly the same power.
One of the examples is shown in Fig. \ref{plot5}.

\begin{figure}[t]
\includegraphics[width=9cm]{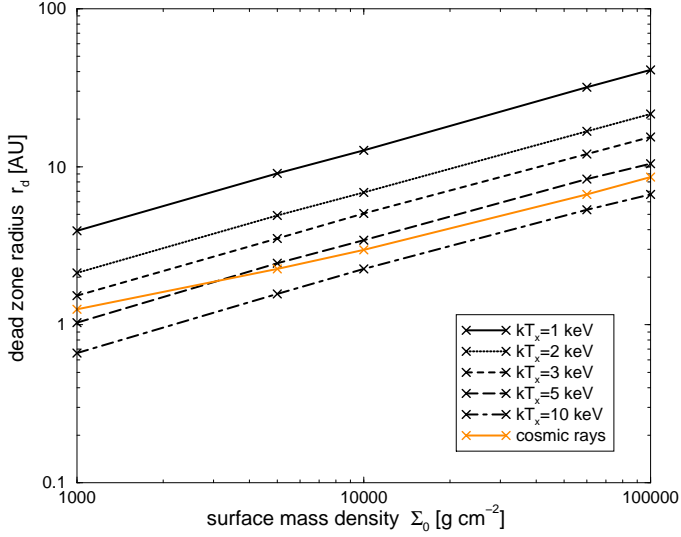}
\caption[Dead zone vs surface mass density 4]{The fiducial dead zone radius as a function of the surface mass density for an X-ray luminosity \(L_{x}=10^{29} 
{\rm erg \ s^{-1}}\). We chose \(R_{eM, crit}=1\) and \(\alpha_{turb}=0.01\) and estimated the dead zone sizes by X-rays alone. Also shown is the dead zone radius estimated by cosmic rays. High energy X-rays dominate cosmic rays in ionizing a disk. \label{plot7}}
\end{figure}

Table \ref{table6} shows the fitting parameters for the dead zone radius by 
X-rays as a function of the surface mass density at 1 AU, \(\Sigma_{0}\).
Again, we can see the similar power for the same choice of the critical magnetic 
Reynolds number \(R_{eM, crit}\) and the alpha parameter \(\alpha_{turb}\). 
One of the examples is shown in Fig. \ref{plot7}.

Table \ref{table7} shows the fitting parameters for the dead zone radius by 
cosmic rays as a function of the surface mass density at 1 AU, \(\Sigma_{0}\).
One of the examples is shown in Fig. \ref{plot7} (the gray solid line with 
crosses). 

\begin{figure}[p]
\includegraphics[width=9cm]{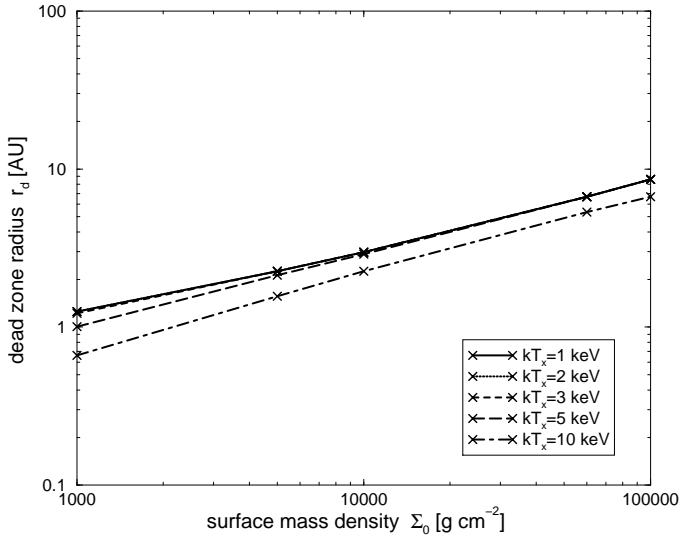}
\caption[Dead zone vs surface mass density 8]{The fiducial dead zone radius as a
function of the surface mass density for an X-ray luminosity \(L_{x}=10^{29} 
{\rm erg \ s^{-1}}\). We chose \(R_{eM, crit}=1\) and \(\alpha_{turb}=0.01\) and estimated the dead zone sizes by total ionization rates.  The dead zone radius changes about an order of magnitude for a range of two orders of magnitude in disk surface mass density. \label{plot8}}
\end{figure}

Table \ref{table8} shows the fitting parameters for the dead zone radius 
estimated by the total ionization rate as a function of the surface mass density at 1 AU, \(\Sigma_{0}\).
Again, we can see the similar power for the same choice of the critical magnetic Reynolds number \(R_{eM, crit}\) and the alpha parameter \(\alpha_{turb}\). 
One of the examples is shown in Fig. \ref{plot8}.
In particular, \(kT_{x}=1 - 3\) keV give a very similar power law relation.
\begin{table}
\caption{The sizes of dead zones as a function of \(kT_{x}\): \(y=ax^{b}\).
The star has X-ray luminosity of $10^{29} {\rm erg \ s^{-1}}$. \label{table5}} 
\begin{center}
\begin{tabular}{|c|c|c|c|c|} \hline
$(R_{eM, crit}, \alpha_{turb})$ & $\Sigma_{0} {\rm [g \ cm^{-2}]}$ & $a$ & $b$ & 
$\chi^2$ \\ \hline \hline
$(1,1)$ & $10^3$ & $2.67$ & $-1.05$ & $0.692e-4$ \\
 & $5\times 10^3$ & $5.97$ & $-0.908$ & $0.583e-3$ \\
 & $10^4$ & $8.47$ & $-0.887$ & $0.140e-2$ \\
 & $6\times 10^4$ & $27.3$ & $-0.887$ & $0.140e-2$ \\
 & $10^5$ & $28.6$ & $-0.883$ & $0.164e-2$ \\ \hline
$(1,0.01)$ & $10^3$ & $3.73$ & $-0.776$ & $0.179e-2$ \\
 & $5\times 10^3$ & $8.58$ & $-0.761$ & $0.201e-2$ \\
 & $10^4$ & $11.9$ & $-0.751$ & $0.213e-2$ \\
 & $6\times 10^4$ & $29.7$ & $-0.772$ & $0.267e-2$ \\
 & $10^5$ & $38.5$ & $-0.786$ & $0.238e-2$ \\ \hline 
$(100,1)$ & $10^3$ & $5.86$ & $-0.609$ & $0.403e-2$ \\
 & $5\times 10^3$ & $12.8$ & $-0.619$ & $0.321e-2$ \\
 & $10^4$ & $18.4$ & $-0.640$ & $0.320e-2$ \\
 & $6\times 10^4$ & $44.8$ & $-0.665$ & $0.370e-2$ \\
 & $10^5$ & $57.9$ & $-0.676$ & $0.296e-2$ \\ \hline
$(100,0.01)$ & $10^3$ & $9.59$ & $-0.458$ & $0.450e-2$ \\
 & $5\times 10^3$ & $20.8$ & $-0.483$ & $0.503e-2$ \\
 & $10^4$ & $29.4$ & $-0.508$ & $0.383e-2$ \\
 & $6\times 10^4$ & $69.7$ & $-0.529$ & $0.387e-2$ \\
 & $10^5$ & $89.5$ & $-0.544$ & $0.423e-2$ \\
\hline
\end{tabular}
\end{center}
\end{table}
\begin{table}
\caption{The sizes of dead zones as a function of \(\Sigma_{0}\): \(y=ax^{b}\). 
The star has the X-ray luminosity of $10^{29} {\rm erg \ s^{-1}}$. 
\label{table6}} 
\begin{center}
\begin{tabular}{|c|c|c|c|c|} \hline
$(R_{eM, crit}, \alpha_{turb})$ & $kT_{x} {\rm [keV]}$ & $a$ & $b$ & $\chi^2$ \\ 
\hline \hline
$(1,1)$ & $1$ & $0.761e-1$ & $0.515$ & $0.315e-4$ \\
 & $2$ & $0.359e-1$ & $0.522$ & $0.443e-3$ \\
 & $3$ & $0.225e-1$ & $0.530$ & $0.658e-3$ \\
 & $5$ & $0.123e-1$ & $0.545$ & $0.220e-2$ \\
 & $10$ & $0.454e-2$ & $0.589$ & $0.820e-2$ \\ \hline
$(1,0.01)$ & $1$ & $0.118$ & $0.508$ & $0.483e-4$ \\
 & $2$ & $0.676e-1$ & $0.501$ & $0.977e-4$ \\
 & $3$ & $0.488e-1$ & $0.500$ & $0.258e-3$ \\
 & $5$ & $0.330e-1$ & $0.502$ & $0.477e-3$ \\
 & $10$ & $0.213e-1$ & $0.501$ & $0.757e-3$ \\ \hline
$(100,1)$ & $1$ & $0.201$ & $0.498$ & $0.496e-4$ \\
 & $2$ & $0.131$ & $0.483$ & $0.178e-5$ \\
 & $3$ & $0.995e-1$ & $0.483$ & $0.178e-5$ \\
 & $5$ & $0.775e-1$ & $0.476$ & $0.474e-4$ \\
 & $10$ & $0.652e-1$ & $0.460$ & $0.324e-4$ \\ \hline
$(100,0.01)$ & $1$ & $0.369$ & $0.483$ & $0.178e-5$ \\
 & $2$ & $0.256$ & $0.471$ & $0.158e-4$ \\
 & $3$ & $0.212$ & $0.467$ & $0.387e-4$ \\
 & $5$ & $0.187$ & $0.455$ & $0.469e-4$ \\
 & $10$ & $0.171$ & $0.441$ & $0.723e-4$ \\ 
\hline
\end{tabular}
\end{center}
\end{table}
\begin{table}
\caption{The sizes of dead zones produced by cosmic rays, as a function of 
\(\Sigma_{0}\): \(y=ax^{b}\). \label{table7}} 
\begin{center}
\begin{tabular}{|c|c|c|c|} \hline
$(R_{eM, crit}, \alpha_{turb})$ & $a$ & $b$ & $\chi^2$ \\ \hline \hline
$(1,1)$ & $0.234e-1$ & $0.485$ & $0.978e-3$ \\
$(1,0.01)$ & $0.650e-1$ & $0.420$ & $0.160e-2$ \\
$(100,1)$ & $0.236$ & $0.341$ & $0.209e-2$ \\
$(100,0.01)$ & $0.830$ & $0.278$ & $0.411e-3$ \\
\hline
\end{tabular}
\end{center}
\end{table}
\begin{table}
\caption{The sizes of dead zones estimated from the total ionization rate as a 
function of \(\Sigma_{0}\): \(y=ax^{b}\). The star has the X-ray luminosity of 
$10^{29} {\rm erg \ s^{-1}}$. \label{table8}} 
\begin{center}
\begin{tabular}{|c|c|c|c|c|} \hline
$(R_{eM, crit}, \alpha_{turb})$ & $kT_{x} {\rm [keV]}$ & $a$ & $b$ & $\chi^2$ \\ 
\hline \hline
$(1,1)$ & $1$ & $0.234e-1$ & $0.485$ & $0.978e-3$ \\
 & $2$ & $0.234e-1$ & $0.485$ & $0.978e-3$ \\
 & $3$ & $0.217e-1$ & $0.492$ & $0.842e-3$ \\
 & $5$ & $0.125e-1$ & $0.539$ & $0.854e-3$ \\
 & $10$ & $0.454e-2$ & $0.589$ & $0.820e-2$ \\ \hline
$(1,0.01)$ & $1$ & $0.650e-1$ & $0.420$ & $0.160e-2$ \\
 & $2$ & $0.650e-1$ & $0.420$ & $0.160e-2$ \\
 & $3$ & $0.617e-1$ & $0.425$ & $0.110e-2$ \\
 & $5$ & $0.403e-1$ & $0.464$ & $0.415e-4$ \\
 & $10$ & $0.213e-1$ & $0.501$ & $0.757e-3$ \\ \hline
$(100,1)$ & $1$ & $0.236$ & $0.341$ & $0.209e-2$ \\
 & $2$ & $0.224$ & $0.346$ & $0.153e-2$ \\
 & $3$ & $0.202$ & $0.356$ & $0.699e-3$ \\
 & $5$ & $0.136$ & $0.391$ & $0.158e-3$ \\
 & $10$ & $0.753e-1$ & $0.439$ & $0.352e-3$ \\ \hline
$(100,0.01)$ & $1$ & $0.830$ & $0.278$ & $0.411e-3$ \\
 & $2$ & $0.730$ & $0.290$ & $0.125e-3$ \\
 & $3$ & $0.591$ & $0.310$ & $0.393e-3$ \\
 & $5$ & $0.443$ & $0.336$ & $0.483e-3$ \\
 & $10$ & $0.278$ & $0.374$ & $0.439e-3$ \\ 
\hline
\end{tabular}
\end{center}
\end{table}
\end{appendix}
\clearpage
\clearpage
\end{document}